\documentclass[showpacs,prb,twocolumn,superscriptaddress]{revtex4}
\usepackage{amsfonts}
\usepackage{graphicx}
\graphicspath{{./.}{./images/}}
\usepackage{amsmath}
\usepackage{amssymb}

\begin{document}

\preprint{UAM, UA, UCM}
\title{Nonadiabatic electron heat pump}
\author{Miguel Rey}
\affiliation{Departamento de F\'{\i}sica Te\'{o}rica de la Materia
Condensada, Universidad Aut\'{o}noma de Madrid, E-28049 Madrid,
Spain} \affiliation{Institut f\"{u}r Physik, Universit\"{a}t
Augsburg, Universit\"{a}tsstra\ss e 1, D-86135 Augsburg, Germany}
\author{Michael Strass}
\affiliation{Institut f\"{u}r Physik, Universit\"{a}t Augsburg,
Universit\"{a}tsstra\ss e 1, D-86135 Augsburg, Germany}
\author{Sigmund Kohler}
\affiliation{Institut f\"{u}r Physik, Universit\"{a}t Augsburg,
Universit\"{a}tsstra\ss e 1, D-86135 Augsburg, Germany}
\author{Peter H\"{a}nggi}
\affiliation{Institut f\"{u}r Physik, Universit\"{a}t Augsburg,
Universit\"{a}tsstra\ss e 1, D-86135 Augsburg, Germany}
\author{Fernando Sols}
\affiliation{Departamento de F\'{\i}sica de Materiales, Universidad
Complutense de Madrid, E-28040 Madrid, Spain} 

\begin{abstract}
We investigate a mechanism for extracting heat from metallic
conductors based on the energy-selective transmission of electrons
through a spatially asymmetric resonant structure subject to ac
driving. This quantum refrigerator can operate at zero net
electronic current as it replaces hot by cold electrons through two
energetically symmetric inelastic channels. We present numerical
results for a specific heterostructure and discuss general trends.
We also explore the conditions under which the cooling rate may
approach the ultimate limit given by the quantum of cooling power.
\end{abstract}

\maketitle

\section{Introduction}

The increasing miniaturization of electronic devices requires a deep
understanding of the generation and flow of heat accompanying
electron motion.\cite{cahi03,giaz06} The quantum of thermal
conductance, which is independent of the carrier
statistics,\cite{rego99} has been recently measured for phonons
\cite{schw00} and photons.\cite{mesc06} A practical and fundamental
issue is the identification of possible cooling mechanisms for
electron systems, a subject less developed than its atom
counterpart.\cite{metc99} Adiabatic electron \cite{hump02,peko06}
and molecular \cite{sega06} pumps may provide reversible heat
engines which would cool with minimum work expenditure. It has also
been proposed that normal-superconductor interfaces can efficiently
cool the normal metal under appropriate conditions of electron
flow.\cite{nahu94,clar05}

\begin{figure}[tbp]
\begin{center}
\includegraphics{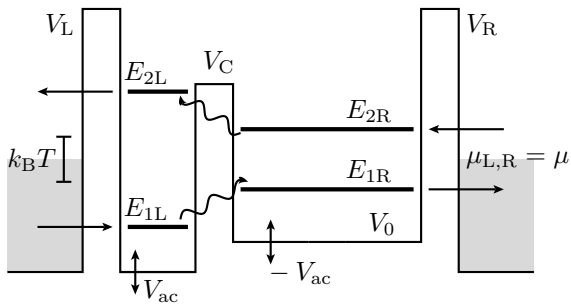} 
\end{center}
\caption{Asymmetric double-well heterostructure used for electron ac
transport calculations. Energy levels are symmetrically placed
around the common Fermi level. Dominant transmission processes
contributing to cooling are shown: in lead R \textit{hot} electrons
are replaced by \textit{cold} electrons, all within a range $\sim
k_{\mathrm{B}}T$ around $\mu$.} \label{eq:model-processes}
\end{figure}

In this paper we explore an alternative electron cooling mechanism
that can operate at zero electric current because it relies on the
idea of replacing hot electrons by cold electrons. The cooling
concept is schematically depicted in Fig.~\ref{eq:model-processes}.
An asymmetric resonant-tunneling structure is formed by two wells
each of which hosts two quasibound states. The four levels are
symmetrically disposed so that the energy difference is smaller in
the right (R) than in the left (L) well. On the other hand, the
difference between the two upper levels is taken to be the same as
that
between the two lower ones, both being equal to the driving frequency: $E_{2%
\text{L}} - E_{2\text{R}} = E_{1\text{R}} -E_{1\text{L}} = \hbar \Omega >0$. In those
conditions, electron transport is dominated by two processes: (i)
electrons in the R electrode with energy $E_{2\text{R}}$ are
inelastically transmitted to the L electrode, where they enter with
energy $E_{2\text{L}} = E_{2\text{R}} + \hbar \Omega$, and (ii)
electrons in the left with energy
$E_{1\text{L}}$ are transmitted to the right while also absorbing a photon.
Unlike in thermionic refrigeration,\cite{maha94,hump05} we may
assume a common chemical potential $\mu = \mu_{\text{L}} =
\mu_{\text{R}}$. Then in the right lead one is
effectively replacing \textit{hot} electrons (with energy $\varepsilon >\mu $%
) by \textit{cold} electrons ($\varepsilon < \mu$). According to
this principle, the right electrode is being \textit{cooled }at the
expense of heating the left electrode. This mechanism, which relies
on the properties of coherent electron transport, may be viewed as
the basis of a quantum refrigerator.\cite{comm-scul03} Under
suitable conditions the two dominant transport mechanisms may cancel
each other yielding a vanishing electric current, which prevents
electrode charging.

\section{Heat pump}

The classification of electrons as hot or cold depending on the
whether its energy is above or below the chemical potential in its
electrode is based on the property that the entropy variation in an
infinitesimal process is given
by $TdS = dU - \mu dN$. For independent electrons, this translates into $%
TdS=(\varepsilon-\mu)dN$, where $\varepsilon$ is the energy of the
electrons being added ($dN>0$) or removed ($dN<0$). In a transport
context, the entropy and temperature variation rates are determined
by the many electron scattering processes continuously taking place
at the interface. We always refer to the equilibrium entropy
eventually reached in the reservoir for the new values of the
conserved quantities energy and particle number.

Since we are ultimately more interested in reducing the temperature
than the entropy, it is important to note that their variations are
not necessarily proportional to each other. One finds
$C_{V}dT = (\varepsilon - \sigma)dN$, where $C_{V}$ is the heat
capacity and $\sigma \equiv \mu -T\left( \partial \mu / \partial
T\right)_{n}$, with $n$ the particle density. In the most
interesting case where the total electron number remains invariant
on average ($\dot{N}=0$), the \textit{total} entropy and temperature
variations are proportional to each other. In the following we
present results for the rate of entropy variation, knowing that it
amounts to temperature variation in the most interesting case of
constant electron number. Specifically, we
compute the heat production rate in lead $\ell =\text{L}, \text{R}$ \cite%
{siva86,mosk02,avro04,arra07}:
\begin{equation}
  \dot{Q}_{\ell } = \sum_{q}(\varepsilon_{q} - \mu_{\ell })
  \dot{N}_{\ell q}\;, \label{eq:Q-dot}
\end{equation}%
$N_{\ell q}$ and $\varepsilon_{q}$ being the electron number and
energy of state $q$ in electrode $\ell$ of chemical potential $\mu_{\ell }=\mu$.

Our goal is to understand the ac thermal transport properties of
quantum-well heterostructures where the electron potential in the
perpendicular $z$ direction has the piecewise constant form shown in
Fig.~\ref{eq:model-processes} while it is uniform in the parallel
$xy$ plane. In such a delocalized system, the independent-electron
approximation is generally adequate. The bottom of the right well
oscillates as $V = V_{0} + V_{\mathrm{ac}}\cos (\Omega t)$ while the
left well operates in phase opposition with the same amplitude and
frequency. To better focus on the main physical aspects, we analyze
first transport through a single channel, later discussing the
effect of many channels.

Electron transport properties can be described in terms of
scattering probabilities. Within a single-channel picture, the
electric current flowing
into lead R under ac driving is given by \cite{wagn99ab}%
\begin{equation}
  \dot{N}_{\text{R}} = \frac{1}{h}\sum_{k=-\infty }^{\infty} \int d \varepsilon %
  \left[ T_{\text{R}\text{L}}^{(k)}(\varepsilon)
  f_{\text{L}}(\varepsilon) - T_{\text{L} \text{R}}^{(k)}
  (\varepsilon) f_{\text{R}}(\varepsilon) \right] \,,
\label{eq:NR-dot}
\end{equation}%
where $f_{\ell }(\varepsilon)$ is the Fermi distribution in lead
$\ell $ and $T_{\ell \ell ^{\prime }}^{(k)}(\varepsilon)$ is the
probability for an electron to be transmitted from lead
$\ell^{\prime}$ to lead $\ell $ while its energy changes from
$\varepsilon $ to $\varepsilon + k\hbar \Omega $, $k$ being an
integer number. Likewise, it can be shown that Eq.~\eqref{eq:Q-dot}
leads to
\begin{align}
  \dot{Q}_{\text{R}} = \frac{1}{h} \sum_{k=-\infty}^{\infty} \int d\varepsilon %
  \big[& (\mu_{\text{R}} - \varepsilon) T_{\text{L}
  \text{R}}^{(k)}(\varepsilon) f_{\text{R}}(\varepsilon) \label{eq:QR-dot} \\
  & + (\varepsilon + k\hbar \Omega - \mu_{\text{R}})
  T_{\text{R}\text{L}}^{(k)} (\varepsilon )f_{\text{L}}(\varepsilon) \notag \\
  & + k\hbar \Omega R_{\text{R}\text{R}}^{(k)} (\varepsilon)
  f_{\text{R}} (\varepsilon) \big] \;, \notag
\end{align}%
where $R_{\text{R} \text{R}}^{(k)}(\varepsilon)$ is the probability
that an electron is reflected in lead R from energy $\varepsilon $
to $\varepsilon + k\hbar \Omega $. Invoking time-reversal symmetry and
the monotonicity of  $f_{\text{R}} (\varepsilon)$, it can be proved that inelastic
reflection always contributes to heating. Therefore any possible
refrigeration of lead
R relying on the transmission scheme depicted in Fig.~\ref%
{eq:model-processes} must be efficient enough to overcome the
heating due to inelastic reflection. The electron scattering
probabilities are calculated exactly following the transfer-matrix
method.\cite{wagn95}
\begin{figure}[tbp]
  \begin{center}
  \includegraphics{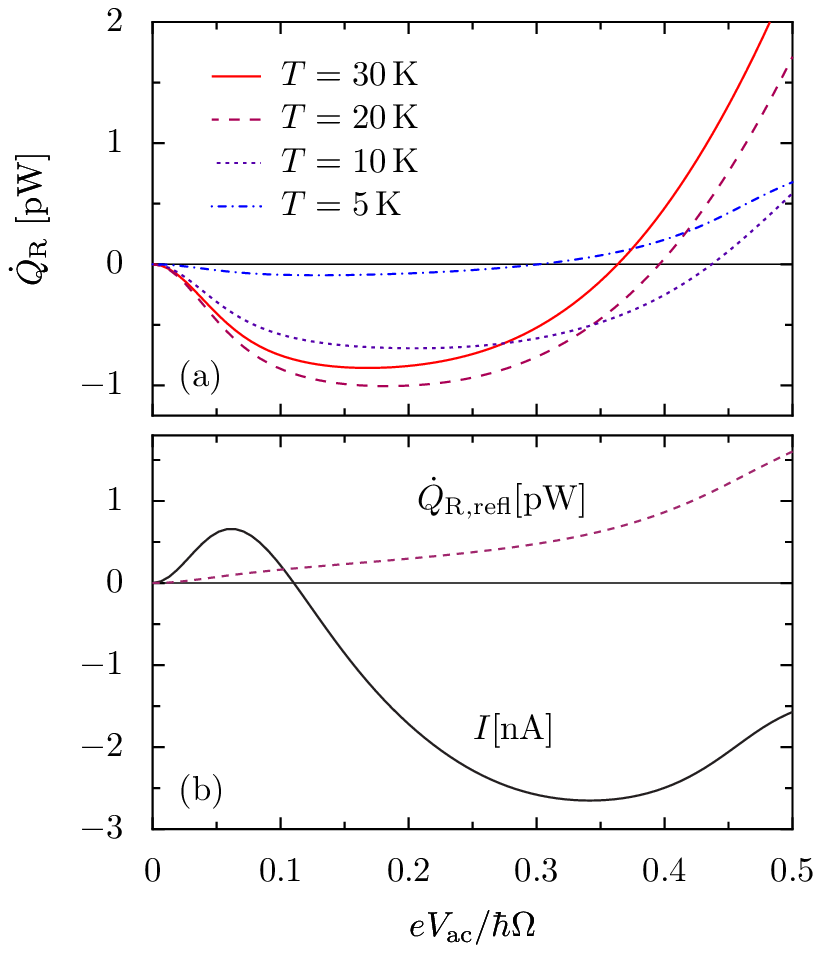}
  \end{center}
\caption{(Color online) (a) Heat production rate in lead R for the
structure of Fig.~\protect \ref{eq:model-processes} as a function of
the driving amplitude $eV_{\mathrm{ac}} $ for various lead
temperatures and $\hbar\Omega=1.94$ meV. See main text for details.
(b) For $T=20\;$K, heating contribution from inelastically reflected
electrons (solid line) and total electric current (line). }
\label{eq:eqqp-vac}
\end{figure}

In Fig.~\ref{eq:eqqp-vac}
we present numerical results for the heat production rate at lead R.
The well lengths are 40 and 80 nm; the heights of the barriers are
$V_{\text{L}} = V_{\text{R}} = 60$ meV and $V_{C} = 30$ meV, measured with
respect to the bottom of the conduction band, and their widths are 4
and 5 nm, respectively; the difference between the bottoms of the
two wells is $V_{0} = 1.5$ meV; the effective electron mass is
$m^{*}=0.07m_{e}$. This results in $E_{2\text{R}}-E_{1\text{R}}=3.4$
meV, as determined e.g. by the dc transmission characteristics. The
structure parameters have been
chosen such that $\hbar \Omega = 1.94$ meV coincides with
$E_{2\text{L}} - E_{2\text{R}}$ and $E_{1\text{R}}-E_{1\text{L}}$. We
take $\mu $ to lie half way between $E_{1\text{R}}$ and
$E_{2\text{R}}$. Clearly, the most negative heat production occurs for
$eV_{\mathrm{ac}} / \hbar \Omega \sim 0.2$. This results from a
combination of nonlinearity, which yields a $V_{\mathrm{ac}}^{2}$
dependence for small $V_{\mathrm{ac}}$, and the
increase of reflection heating (see Fig.~\ref{eq:eqqp-vac}b)
reinforced by the suppression of electron transmission through the
dominant single-photon channels as $eV_{\mathrm{ac}}/\hbar \Omega $
approaches the first zero of
the first-order Bessel function.\cite{rey05} The result is that
$|\dot{Q}_{\text{R}}|$ goes through a maximum for a moderate value of
$eV_{\mathrm{ac}}/\hbar \Omega$.

Another interesting feature is that, as a function of $T$, the
cooling rate
is maximized for $T \sim 20$ K, which is roughly $(E_{2\text{R}} -
E_{1\text{R}})/2$. If the temperature is too low, the level 2R is empty and 1R
is full, which inhibits the exchange of electrons. If it is too
high, the cooling rate saturates as $T_{\text{R}}$ increases, and
even decreases slightly because $T_{\text{L}}$ (here equal to
$T_{\text{R}}$) also increases. Later we
argue more generally that cooling is optimized when, not only
$(E_{2\text{R}} - E_{1\text{R}})/2$ but also $\Gamma/2$ (the half-width of the
transmitting channels) is of order $k_{\mathrm{B}} T_{\text{R}}$.
Here $\Gamma/2 \sim 0.2$ meV, noticeably smaller than
$k_{\mathrm{B}} T_{\text{R}}$.

\begin{figure}[tbp]
\begin{center}
\includegraphics{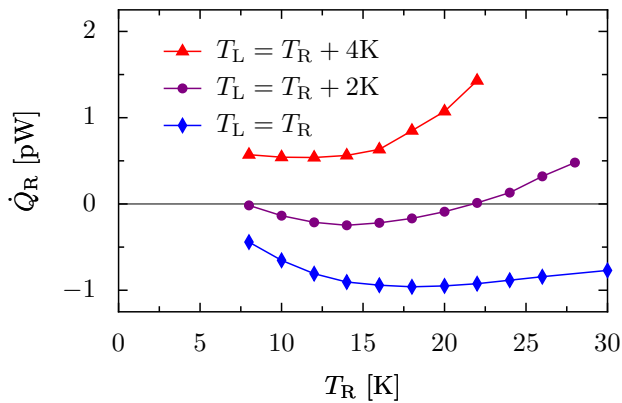}
\end{center}
\caption{(Color online) Cooling rate of the R electrode as a
function of its temperature $T_{\text{R}}$ for several values of
$T_{\text{L}} \geq T_{\text{R}}$, with $\mu$ adjusted to yield zero
electric current, for $\hbar\Omega=1.94$ meV and
$eV_{\mathrm{ac}}/\hbar\Omega=0.2$.} \label{eq:cooling-at-zero-curr}
\end{figure}

The potentially most interesting scenario is that where cooling
takes place while the net electric current is zero (in a classical
context see Ref. \onlinecite{mara07}). That this is not generally
the case can be inferred from the inset of Fig.~\ref{eq:eqqp-vac}.
If we fix the structure and driving parameters, then the chemical
potential and the temperatures are left as the independent
variables. If $\mu $ is adjusted to
satisfy the constraint $\dot{N}_{\text{R}}=0$, the cooling rate
$\dot{Q}_{\text{R}}$ becomes a unique function of $T_{\text{L}}$ and
$T_{\text{R}}$. Figure ~\ref{eq:cooling-at-zero-curr} shows the
resulting cooling rate as a function of $T_{\text{R}}$ for several
values of $T_{\text{L}}$. Remarkably, we
observe that the heat production in R can be negative even for
$T_{\text{L}} > T_{\text{R}}$. We conclude that it is technically possible to extract
heat from the cold reservoir and pump it to the hot reservoir with a
vanishing net electric current. Thermodynamically, such a
refrigeration process requires external work, which here amounts to
$2\hbar \Omega $ per useful scattering event and is provided by the
classical ac source. In practice, inelastic reflection will further
reduce the efficiency.

In a 3D context one must generalize Eq. (3) to include a sum over
transverse modes while replacing $\mu_{\text{R}}$ by
$\mu_{\text{R}}-\hbar^2 \kappa^2/2m^{*}$, where each channel is
characterized by its parallel wave vector $\vec{\kappa}$.
For fixed $T_{\text{L}}$ and $T_{\text{R}}$, $\dot{Q}_{\text{R}}$ remains
negative within a finite range of $\mu $ values (not shown). This
suggests that, after summing the contributions from the many
transverse channels, global cooling is still possible in a suitably
designed 3D interface.

\section{The quantum limit}

Once we have proved that it is in principle possible to pump heat
from a cold to a hot reservoir by coherent control of electron
transmission, it is natural to ask whether there is any fundamental
limit to the maximum cooling rate per quantum channel which would
play a role analogous to the quantum of
electric or thermal conductance ($e^{2}/h$ and $\pi^{2}
k_{\mathrm{B}}^{2}T/3h$, respectively). It seems evident that the maximum cooling
rate should be achieved in an ideal setup where a metal at
temperature $T$ is connected through a totally transparent interface
to another metal at the same chemical potential but at zero
temperature. The result is the quantum of cooling power:
\begin{equation}
  C_{Q} \equiv |\dot{Q}|_{\text{max}} = \frac{2}{h} \int_{0}^{\infty}
  d \varepsilon \, \varepsilon \, f(\varepsilon) = 
  \frac{\pi^{2}}{6}\frac{k_{\mathrm{B}}^{2} T^{2}}{h}\;, \label{eq:q-limit}
\end{equation}%
where $f(\varepsilon) \equiv \lbrack \exp (\varepsilon /
k_{\mathrm{B}}T) + 1]^{-1}$ and $\pi^{2} k_{\mathrm{B}}^{2}/6h=473$ fW K$^{-2}$.
Following information theory arguments, a similar result can be
derived \cite{pend83,blen00}. Differentiation of (\ref{eq:q-limit}) yields the
quantum of thermal conductance. Invoking only time-reversal symmetry
and unitarity, Eq.
(\ref{eq:QR-dot}) can be shown to satisfy (with $T_{\text{R}}=T$)%
\begin{equation}
  \dot{Q}_{\text{R}}\geq -C_{Q}~,  \label{inequality}
\end{equation}%
for arbitrary electrodes (including $\mu_{\text{L}} \neq
\mu_{\text{R}}$) and driving parameters, thus confirming that $C_{Q}$
is an upper bound to the cooling rate.

The quantum limit may be intuitively understood as follows:
$k_{\mathrm{B}}T$ is the maximum amount of heat that can be carried
away in an elementary process. Such processes take place at a rate
$\sim|\dot{Q}|/k_{\mathrm{B}}T$, which cannot exceed
$h/k_{\mathrm{B}}T$ if one is to avoid effective
heating caused by energy uncertainty. This results in $|\dot{Q}|
\lesssim k_{\mathrm{B}}^{2}T^{2}/h$, as given more precisely in
\eqref{eq:q-limit}. This argument suggests that
$k_{\mathrm{B}}^{2}T^{2}/h$ is also a quantum limit for the cooling
rate per active degree of freedom (with characteristic energy scale
$\ll k_{\mathrm{B}}T$) when cooling acts on the volume instead of
through the surface, as e.g. in laser cooling.\cite{metc99,comm1}

The question naturally arises of whether in a setup like that of Fig.~\ref%
{eq:model-processes} it is possible to approach the quantum limit.
This problem can be explored analytically within a simple model. We
neglect reflection heating and assume that electron transmission is
dominated by two one-photon inelastic channels, or pipelines
\cite{wagn99ab} (see Fig.~\ref{eq:model-processes}), named
$\tau_{u}(\varepsilon )\equiv T_{\text{L} \text{R}}^{(1)} (\varepsilon
)$ and $\tau _{d}(\varepsilon )\equiv T_{\text{L} \text{R}}^{(-1)}
(\varepsilon) = T_{\text{R}\text{L}}^{(1)}(\varepsilon -\hbar
\Omega)$, which peak at energies $E_{2\text{R}}$ and
$E_{1\text{R}}$, respectively, always satisfying the unitarity
requirement $\tau_{u} + \tau_{d}<1$. We take the energy origin at
the middle point $(E_{1\text{R}} + E_{2\text{R}})/2$, so that
$E_{2\text{R}} = -E_{1\text{R}} \equiv \varepsilon_{0}>0$. Electrons
entering the scattering region from R with initial energy $\pm
\varepsilon_{0}$ will be transmitted with final energy
$\varepsilon_{0} \pm \hbar \Omega $ through the upper (lower) channel.
If we assume the pipelines to be symmetric, $\tau_{u}(\varepsilon) =
\tau_{d} (-\varepsilon) \equiv \tau (\varepsilon)$, and $\mu=0$, we
obtain $\dot{N}_{\text{R}}=0$ and
\begin{equation}
  \dot{Q}_{\text{R}} = -\frac{2}{h} \int d \varepsilon \, \varepsilon
  ~ [f_{\text{R}} (\varepsilon) - f_{\text{L}} (\varepsilon + \hbar \Omega)]\tau
  (\varepsilon)~. \label{eq:QR-symmetric}
\end{equation}%
We note that at zero temperature Eq.~\eqref{eq:QR-symmetric} only
yields heating, as should be expected.

If we take $\tau (\varepsilon)$ to be a Lorentzian of width $\Gamma$
centered around $\varepsilon_{0}$, some complications arise due
to its slow decay for large $|\varepsilon - \varepsilon_{0}|$. For
instance, for large enough $\Omega $, we always find heating
$\dot{Q}_{\text{R}}\propto \ln \Omega $. On the other hand, for
small $\Omega $, $\dot{Q}_{\text{R}}<0$ if and only if $T_{\text{L}} <
T_{\text{R}}$. We conclude that, in the interesting case $T_{\text{L}}
> T_{\text{R}}$, cooling of the R electrode can only occur
within a finite range of $\Omega $ values. This range shrinks to zero
for $T_{\text{L}}$ large.

An interesting question is whether, given two electrodes with
$T_{\text{L}} > T_{\text{R}}$, it is always possible to design an ac resonance
structure
yielding $\dot{Q}_{\text{R}}<0$, and whether $|\dot{Q}_{\text{R}}|$
can ever approach the quantum limit. In
Eq.~\eqref{eq:QR-symmetric}, $g(\varepsilon )\equiv \varepsilon ~[f_{\text{R}%
}(\varepsilon )-f_{\text{L}}(\varepsilon +\hbar \Omega )]>0$ only in the
interval
$0<\varepsilon <\bar{\varepsilon} \equiv \hbar \Omega T_{\text{R}}/(T_{\text{L} %
}-T_{\text{R}})$. For $T_{\text{L} }\rightarrow T_{\text{R}}$, we have $\bar{%
\varepsilon}\rightarrow \infty $; however, the integrand decays
exponentially on a scale $\sim k_{\mathrm{B}}T$ after having peaked at $%
\varepsilon \sim k_{\mathrm{B}}T$. Therefore, cooling comes
effectively from the interval $0<\varepsilon < \varepsilon_{1}$,
where $\varepsilon_{1} \equiv \min \{\bar{\varepsilon},2k_{\mathrm{B}}
\bar{T}\}$ and $\bar{T} \equiv (T_{\text{L} }+T_{\text{R}})/2$. To
potentiate the contribution from that segment, we may design $\tau
(\varepsilon)$ to be centered at $\varepsilon_{0} \simeq
\varepsilon_{1}/2$. If $\Gamma \rightarrow 0$, $\dot{Q}_{\text{R}}$ is
guaranteed to become negative, although with a vanishing magnitude
$|\dot{Q}_{\text{R}}| \propto \Gamma $. A typical optimal value is
$\Gamma \sim \varepsilon_{1}$. We conclude that the cooling rate is
maximized for $\varepsilon_{0} \sim \Gamma /2 \sim \varepsilon_{1}/2$.
The peak at $k_{\mathrm{B}} T \sim \varepsilon_{0}$ for $T_{\L} =
T_{\text{R}}$ is confirmed by the lowest curve of Fig.~\ref{eq:cooling-at-zero-curr}.

If $\tau (\varepsilon)$ decays sufficiently fast away from the
region where $g(\varepsilon)>0$, one may estimate
$|\dot{Q}_{\text{R}}| \sim - (2/h)\varepsilon_{1} \tau_{\max}
g_{\max}$. For $T_{\text{L}} \rightarrow T_{\text{R}}$ we have both
$\varepsilon_{1}$ and $g_{\max}$ of order $k_{\mathrm{B}}T$, assuming
$\hbar \Omega \gg k_{\mathrm{B}}T$. In those conditions,
$|\dot{Q}_{\text{R}}| \sim C_{Q}$ provided $\tau_{\max}$ is
close to unity. By contrast, the cooling rate cannot approach
the quantum limit if $T_{\text{L}}$ grows substantially above
$T_{\text{R}}$ or if $\tau (\varepsilon)$ decays slowly, like in a
Lorentzian resonance, since then the contribution from
$g(\varepsilon)<0$ cannot be neglected.

Careful inspection of Eq.~\eqref{eq:QR-symmetric} reveals that
$\dot{Q}_{\text{R}}$ increases monotonically as $T_{\text{R}}$ decreases.
Thus, if we start cooling the R electrode, $\dot{Q}_{\text{R}}$
begins to increase until it eventually becomes zero. At that point,
no further cooling is possible. We have reached the lowest possible
temperature for the refrigeration process defined by $\tau
(\varepsilon )$ and $\Omega $. We are limited by the lack of
sufficient energy resolution: when $k_{\mathrm{B}}T_{\text{R}}$
becomes small compared with the linewidth $\Gamma $, no heat pumping
is possible for $T_{\text{L}}>T_{\text{R}}$.

\section{Discussion}

The quantum refrigerator which we have investigated may be viewed as
a realization of Maxwell's demon \cite{maxw1871} as it selectively
lets hot electrons out while it only lets cold electrons in. The
required work is provided by the external ac source which, combined
with the spatial asymmetry of the structure, rectifies electron
motion. The work might also be extracted from a hot ohmic
resistor.\cite{peko07} Alternative schemes to that of
Fig.~\ref{eq:model-processes} are of course possible: One may design
two superlattices each of them having two narrow bandwidths yielding
a similar level distribution. A potential advantage of such a device
would be that, away from resonance, transmission would decay fast.
Thus it would show interesting features such as cooling for
arbitrary large $\Omega$ and, as discussed above, the guaranteed
existence of a driving structure that brings the cooling rate close
to the quantum limit. A non-resonant mechanical mismatch at the
interface would prevent phonons from short-circuiting electron
cooling during operation close to such a limit.

In conclusion, we have identified a mechanism for nonadiabatically
pumping heat from a cold to a hot electron reservoir which is based
on the coherent control of electron ac transport and which can
operate at zero average electric current. On the basis of electron
transport considerations, the quantum of cooling power $C_{Q}$ has
been shown to be an upper bound to the cooling rate per quantum
channel. We have investigated the case of Lorentzian resonances,
where approaching the quantum limit is generally not feasible. We
have noted however that with sharper resonances it is always
possible to design a driven interface that provides cooling at a
rate close to the quantum limit.

\begin{acknowledgments}
This research has been supported by MEC (Spain) No. FIS2004-05120,
the EU Marie Curie RTN Programme No. MRTN-CT-2003-504574, the
Ram\'{o}n Areces Foundation, DFG (Germany) through SPP 1243 No.
HA1517/30-1, and the Spain-Germany Programme of Acciones Integradas
(DAAD and MEC). Support from the Centro de Computaci\'{o}n
Cient\'{\i}fica (UAM) is also acknowledged. SK and PH gratefully
acknowledge support by the German Excellence Initiative via the
``Nanosystems Initiative Munich (NIM)''.
\end{acknowledgments}


\begin{thebibliography}{99}
\bibitem{comm1} In fact, the existence of a minimum temperature in laser
cooling can be viewed as a consequence of the cooling rate having to
stay below the quantum limit.
\end{thebibliography}
\end{document}